\documentclass[journal]{IEEEtran}
\IEEEoverridecommandlockouts
\usepackage{amsmath,amsfonts}
\usepackage{mathtools}
\usepackage{amsthm}
\usepackage{array}
\usepackage{textcomp}
\usepackage{stfloats}
\usepackage{url}
\usepackage{microtype}
\usepackage{verbatim}
\usepackage{graphicx}
\usepackage{cite}
\usepackage[colorlinks=true, linkcolor=blue, citecolor=blue, urlcolor=blue]{hyperref}
\usepackage{enumitem}

\usepackage[linesnumbered,lined,algoruled,commentsnumbered]{algorithm2e}

\usepackage{caption}
\usepackage{subcaption}
\usepackage{stfloats}
\usepackage{lipsum} % for placeholder text
\usepackage[font=small,skip=0pt]{caption}
\usepackage[export]{adjustbox}
\usepackage[inkscapelatex=false]{svg}

\SetCommentSty{mycommfont}
\theoremstyle{plain}

\usepackage{units}
\newcommand{\nth}[1]{{#1}^{\text{th}}}

\newcommand{\RD}[0]{r_{\mathrm{\scalebox{0.6} {RD} }}}
\newcommand{\MRD}[0]{r_{\mathrm{\scalebox{0.6} {EMRD} }}}
\newcommand{\MSD}[0]{r_{\mathrm{\scalebox{0.6} {MSMD} }}}

\usepackage{acronym}
\input{acronym}
\begin{document}
\title{Spatial Degrees of Freedom in Near Field MIMO: Experimental Validation of Beamspace Perspective
}

\author{Ahmed~Hussain,
        Asmaa~Abdallah,~\IEEEmembership{Senior Member,~IEEE}, 
        Ahmed~Nasser,~\IEEEmembership{Senior Member,~IEEE}, 
        Abdulkadir~Celik,~\IEEEmembership{Senior Member,~IEEE}, and~Ahmed~M.~Eltawil,~\IEEEmembership{Senior Member,~IEEE}%
\thanks{Ahmed Hussain, Asmaa Abdallah, Ahmed Nasser, and Ahmed M. Eltawil are with the Computer, Electrical, and Mathematical Sciences and Engineering (CEMSE) Division, King Abdullah University of Science and Technology (KAUST), Thuwal 23955-6900, Saudi Arabia. Abdulkadir Celik is with School of Electronics and Computer Science, University of Southampton, SO17 1BJ UK.}
}
\maketitle

\begin{abstract}
Conventional far-field multiple-input multiple-output (MIMO) channels are limited to a single spatial degree of freedom (DoF) under a line-of-sight (LoS) condition. In contrast, the radiative near field (NF) supports multiple spatial DoF, enabled by spherical wavefronts and the reduced spatial footprint at short ranges. While recent research indicates that the effective DoF (EDoF) increases in NF, experimental validation and clear identification of the transition distances remain limited. In this letter, we develop an intuitive framework for characterizing the EDoF of a ULA-based MIMO system and derive two complementary analytical expressions: a closed-form formulation that relates the EDoF to the physical transmit beamwidth and receive aperture, and a discrete formulation based on the discrete Fourier transform (DFT) domain angular decomposition of the NF spherical wavefront, which is well suited for experimental evaluation. We further introduce the effective MIMO Rayleigh distance (EMRD) and the maximum spatial multiplexing distance (MSMD), which mark the distances where the EDoF reduces to one and attains its maximum, respectively. Experimental measurements using widely spaced phased arrays closely match the theoretical EDoF trends and validate the proposed distance metrics.

%In this letter, we develop an intuitive framework for characterizing the EDoF of ULA-based MIMO system and derive two analytical expressions: one relating the EDoF to the transmit beamwidth and receive aperture, and another based on the DFT-domain NF angular spread. We further introduce the effective MIMO Rayleigh distance (EMRD) and the maximum spatial multiplexing distance (MSMD), which mark the distances where the EDoF reduces to one and attains its maximum, respectively. Experimental measurements using distributed phased arrays closely match the theoretical EDoF trends and validate the proposed distance metrics.
\end{abstract}

\section{Introduction} \label{sec_I}
\IEEEPARstart{F}{uture} wireless networks employing large antenna arrays at high frequencies are expected to operate in the radiative \ac{NF}, where spherical wavefronts enhance both single-user and multi-user capacity \cite{11095387}. In the \ac{NF}, the finite beamdepth enables spatial separation of \acp{UE} along the range dimension, thereby improving multiuser capacity \cite{10934779}. In the \ac{FF}, a \ac{LoS} \ac{MIMO} channel is typically rank one, supporting only a single data stream. In contrast, the radiative \ac{NF}, with its inherent spherical wavefronts, offers enriched spatial \ac{DoF} \cite{ouyang2023near}, enabling multiple data streams even in \ac{LoS} case.

\begin{figure}[t]
\centering
    \includegraphics[width=.9\columnwidth]{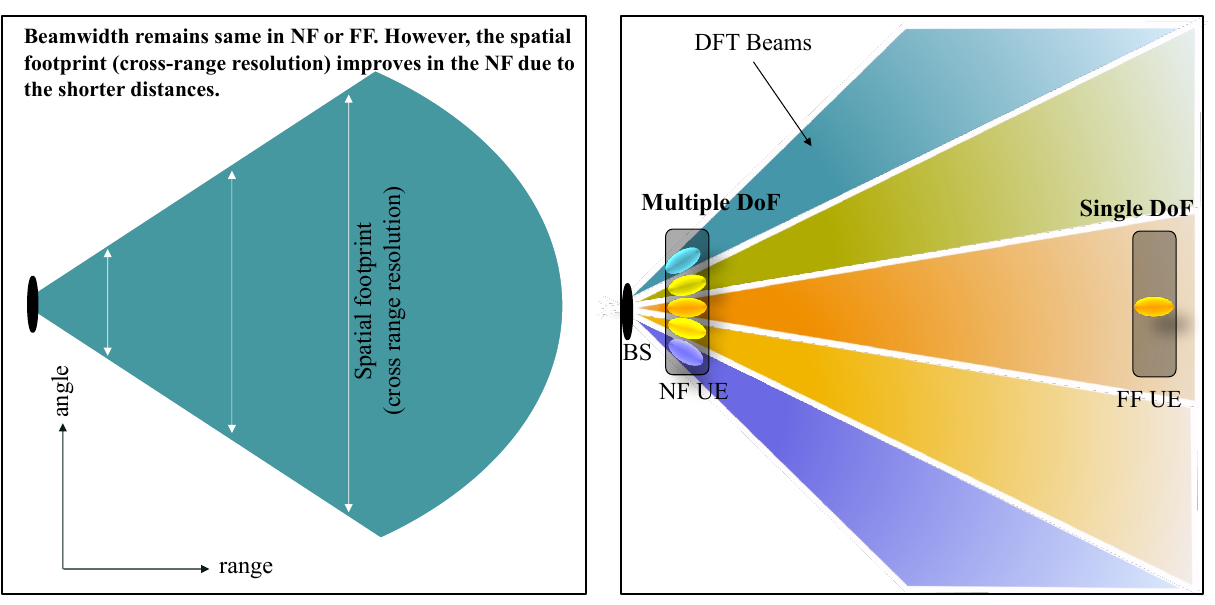} % or inkscapearea=bbox
\setlength{\belowcaptionskip}{-22pt}
\caption{Multiple EDoF in \ac{NF} communication.}
\label{fig1_abstract}
\end{figure}

Spatial \ac{DoF} represent the parallel communication modes determined by the propagation environment and array geometry~\cite{10934778}. While the array geometry imposes an upper bound on the achievable \ac{DoF}, the usable \ac{DoF} are limited by the minimum of the transmit and receive antenna counts. The spatial \ac{DoF} also vary with the angular extent of the radiated field, set by the transmit beamwidth, and the physical size of the receiving aperture. Although the angular beamwidth is invariant with distance, the \emph{spatial footprint} expands linearly with range as shown in Fig.~\ref{fig1_abstract} (left). Consequently, capturing multiple angular beams in the \ac{FF} requires prohibitively large receive apertures; for instance, spatial footprint corresponding to a beamwidth of $3^\circ$ spans $0.52\,\mathrm{m}$ at a distance of $10\,\mathrm{m}$, but expands to $52\,\mathrm{m}$ at $1\,\mathrm{km}$. Furthermore, beamfocusing-based \ac{NF} beams inherently comprise multiple angular components due to the spherical wavefront being a superposition of plane waves \cite{11095387}. As illustrated in Fig.~\ref{fig1_abstract} (right), the reduced spatial footprint and inherent curvature of spherical wavefronts in the \ac{NF} may potentially unlock higher spatial \ac{DoF}. 

The spatial \ac{DoF} are quantified through the singular values of the \ac{MIMO} channel, which typically exhibit a step-like distribution, indicating that the corresponding subchannels are not of equal quality \cite{ahmed2025NF}. Only the singular values exceeding a certain threshold effectively contribute to capacity, giving rise to the concept of \ac{EDoF}. In the \ac{NF}, higher \ac{EDoF} can be achieved either by enlarging the aperture or by increasing the carrier frequency. The latter, however, introduces severe coverage limitations, whereas expanding the aperture requires a very large number of antennas, leading to prohibitive hardware and signal-processing complexity. Although theory suggests that the \ac{EDoF} effectively reduces to one in the \ac{FF} and increases in the \ac{NF}~\cite{10934778}, experimental validation remains limited. Moreover, the exact distance at which the \ac{EDoF} first exceeds one, and the distance at which it reaches its maximum, has not been clearly established.

To address these gaps, we develop an intuitive framework for characterizing the \ac{EDoF} and validate it through experimental measurements. Specifically, we derive two complementary analytical expressions for the \ac{EDoF} in \ac{ULA}-based \ac{MIMO} systems. The first is a closed-form expression that relates the \ac{EDoF} to the physical transmit beamwidth and receive aperture, providing insight into its distance-dependent behavior. The second is a discrete formulation based on \ac{DFT}, which captures the angular decomposition of the spherical wavefront. We further introduce the \ac{EMRD}, defined as the distance at which the \ac{EDoF} reduces to one, and the \ac{MSD}, which denotes the distance at which the maximum \ac{EDoF} is attained. Experimental measurements closely follow the theoretical analysis and validate the proposed distance metrics.
%To address these gaps, we develop an intuitive framework for understanding the \ac{EDoF} and validate it through measurements. We derive two analytical expressions for the \ac{EDoF} in \ac{ULA}-based \ac{MIMO} systems: one relating the \ac{EDoF} to the transmit beamwidth and receive aperture, and another based on a \ac{DFT}-domain characterization of the \ac{NF} angular spread, which governs the singular-value distribution and thus the channel capacity. We further introduce the \ac{EMRD}, defined as the distance at which the \ac{EDoF} reduces to one, and the \ac{MSD}, which marks the distance at which the maximum \ac{EDoF} is achieved. Experimental measurements closely follow the theoretical analysis and the proposed distance metrics. 
  
\section{System Model} \label{Sec_II}
\begin{figure}[t]
\centering
    \includegraphics[width=.8\linewidth]{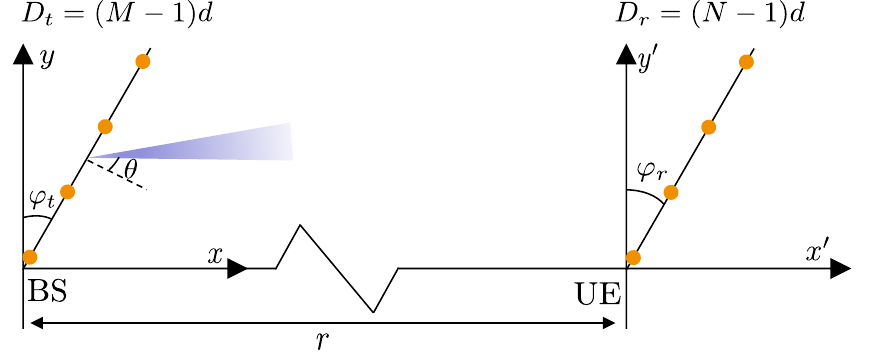} % or inkscapearea=bbox
\setlength{\belowcaptionskip}{-20pt}
\setlength{\abovecaptionskip}{0pt}
\caption{A \ac{MIMO} system model with \ac{ULA} at the \ac{BS} and \ac{UE}.}
\label{fig2_system_model}
\end{figure}

We consider a point-to-point \ac{LoS} \ac{MIMO} system, as illustrated in Fig.~\ref{fig2_system_model}. The \ac{BS} and \ac{UE} are equipped with \acp{ULA} comprising $M$ transmit and $N$ receive antennas, respectively. The inter-element spacing is set to $d=\lambda/2$, yielding approximate aperture lengths of $D_t \approx Md$ at the \ac{BS} and $D_r \approx Nd$ at the \ac{UE}. The \ac{UE} is located at a distance $r$ from the \ac{BS}. The \acp{ULA} at the \ac{BS} and \ac{UE} are oriented at angles $\varphi_t$ and $\varphi_r$, respectively, with respect to their local coordinate systems. The channel coefficient of the \ac{MIMO} channel $\mathbf{H} \in \mathbb{C}^{M \times N}$ between the $\nth{m}$ transmit and $\nth{n}$ receive antenna elements is given by
\begin{equation}
h_{m,n} = \sqrt{g_{m,n}}\, e^{-j\tfrac{2\pi}{\lambda}(r_{m,n}-r)},
\label{eqn_A1}
\end{equation}
where $g_{m,n}= \tfrac{\lambda^2}{(4\pi)^2 (r_{m,n})^2},$ denotes the free-space pathloss and $r_{m,n}$ denote the distance between the $\nth{m}$ transmit and $\nth{n}$ receive elements, approximated as
{\small
\begin{align}
&r_{m,n}  
\!=\! \sqrt{\big(r + m d \sin\varphi_t \!- \!n d \sin\varphi_r\big)^2
       \!+\! \big(m d \cos\varphi_t - n d \cos\varphi_r\big)^2}, \nonumber\\
&\!\approx r
       + m d \sin\varphi_t
       - n d \sin\varphi_r
       + \frac{\big(m d \cos\varphi_t - n d \cos\varphi_r\big)^2}{2r}.
\label{eqn_A2}
\end{align}}
\begin{comment}
where the second expression follows from a first-order Taylor expansion. Accordingly, the $\nth{(m,n)}$ element of the \ac{NF} array response vector $b_{m,n} = e^{-j\nu(r_{m,n}-r)}$ in \eqref{eqn_A1} can be written as
\begin{align}
b_{m,n} \approx 
    e^{-j\nu \left( 
        m d \sin\varphi_t
        - n d \sin\varphi_r
        + \tfrac{(m d \cos\varphi_t - n d \cos\varphi_r)^2}{2r}
    \right)},
\label{eqn_A3}
\end{align}
where $\nu = 2\pi / \lambda$ is the wavenumber. The far-field counterpart of \eqref{eqn_A3} is obtained by neglecting the quadratic phase term term as
\begin{equation}
a_{m,n} = e^{-j\nu\left( m d \sin\varphi_t
       - n d \sin\varphi_r\right)}.
\label{eqn_A4}
\end{equation}
\end{comment}
Given a \ac{MIMO} channel $\mathbf{H} \in \mathbb{C}^{M \times N}$, the spatial \ac{DoF} represent the number of independent sub-channels that support parallel transmission modes. Mathematically, this corresponds to the number of non-zero singular values obtained from the \ac{SVD} of $\mathbf{H}$, or equivalently, the rank of the Gramian matrix $\mathbf{R} = \mathbf{H}^{\mathrm{H}}\mathbf{H}$. While the channel rank determines the potential number of parallel streams, the actual channel capacity also depends on the distribution of eigenvalues. Hence, the \ac{EDoF} is often used to quantify the number of effective parallel channels at high \ac{SNR}. The eigenvalues $\mu_i$ exhibit step-like behavior, remaining approximately equal up to a certain index $\mathsf{EDoF}$, beyond which they decay sharply to zero. Assuming $\mathsf{EDoF}$ equal eigenvalues $(\mu_1 = \mu_2 = \cdots = \mu_\mathsf{EDoF} = \mu)$, the sum of all eigenvalues equals the trace of the Gramian $\sum_{i=1}^{\min(M,N)} \mu_i = \operatorname{tr}(\mathbf{H} \mathbf{H}^{\mathsf{H}}) = MN$. Thus, with only $\mathsf{EDoF}$ non-zero eigenvalues, we have
\begin{equation}
\mathsf{EDoF} \mu = MN \quad \Rightarrow \quad \mu = \tfrac{MN}{\mathsf{EDoF}}.
\label{eqn_A5}
\end{equation}
Diagonalizing $\mathbf{H}$ yields $\mathsf{EDoF}$ parallel channels, each with an equal gain $\mu$. With equal eigenvalues, the optimal power allocation is uniform, i.e., $P_i = P/\mathsf{EDoF}$, leading to the single-user capacity expression as
\small
\begin{align}
C &= \sum_{i=1}^{\mathsf{EDoF}}\log_2\!\left(1+\frac{P_i \mu}{\sigma^2}\right)
   = \mathsf{EDoF} \log_2\!\left(1+\frac{P\mu}{\mathsf{EDoF}\sigma^2}\right),
   \label{eqn_A6}\\
  C&= \mathsf{EDoF} \log_2\!\left(1+\frac{\rho}{\mathsf{EDoF}^2}\right), 
   \qquad \rho = \frac{PMN}{\sigma^2},
   \label{eqn_A7}
\end{align}\normalsize
The final simplified form is obtained by substituting $\mu$ from \eqref{eqn_A5}, and $\sigma^2$ denotes the noise power. At high SNR, the capacity scales approximately as $C \approx\mathsf{EDoF}\log_2(\rho/\mathsf{EDoF}^2)$, increasing almost linearly with $\mathsf{EDoF}$.  

\section{Effective Spatial Degrees of Freedom} \label{sec-III}
As illustrated in Fig.~\ref{fig1_abstract}, the reduced spatial footprint and the spherical wavefronts in the \ac{NF} enable higher spatial \ac{DoF}. In this section, we derive two complementary analytical expressions for the \ac{EDoF}. The first is a closed-form expression based on the physical beamwidth, which varies continuously with distance and provides theoretical insight into the spatial scaling behavior. The second is based on an angular decomposition of the spherical wavefront, leading to a discrete formulation that is more suitable for experimental evaluation.
\subsection{Electromagnetic Perspective}
The spatial \ac{DoF} denote the number of linearly independent columns in the channel matrix $\mathbf{H}$, whereas the \ac{EDoF} represent the subset of mutually orthogonal columns. From an electromagnetic standpoint, the \ac{EDoF} represents the number of transmit beams that remain distinguishable at the receiver. It can be approximated as the ratio between the receive aperture length $D_r$ and the spatial footprint of the transmit beam \cite{ahmed2025NF}. This ratio indicates how many transmit beams illuminate the receiver aperture with minimal mutual interference. The spatial footprint referred to as the cross-range resolution $\Delta_{\mathrm{CR}}$ is determined by the transmit beamwidth $\theta_{3\text{dB}}$. For a \ac{ULA} with transmit aperture $D_t$, the cross-range resolution and beamwidth are 
\begin{equation}
  \Delta_{\mathrm{CR}} = r\,\theta_{3\text{dB}} \quad \text{and} 
  \quad  
  \theta_{3\text{dB}} \approx \frac{\lambda}{D_t \cos\varphi_t},
  \label{eqn_BW}
\end{equation}
respectively, leading to the following closed-form expression for the \ac{EDoF}:
\begin{equation}
    \mathsf{EDoF}_1 \approx \frac{D_r\cos\varphi_r}{\Delta_{\mathrm{CR}}}
    = \frac{D_r D_t}{\lambda r}\cos\varphi_t \cos\varphi_r.
    \label{eqn_EDoF1}
\end{equation}
Although the angular beamwidth $\theta_{3\text{dB}}$ remains invariant across the \ac{NF} and \ac{FF}, the cross-range resolution $\Delta_{\mathrm{CR}}$ improves as the transmitter–receiver separation $r$ decreases. As a result, the \ac{NF} yields a higher \ac{EDoF} due to its finer cross-range discrimination. 

Beyond the reduced spatial footprint, spherical wave propagation further enhances the \ac{EDoF}. Unlike planar wavefronts, which induce a linear phase profile across the array and support a single dominant spatial mode, spherical wavefronts introduce phase curvature across the aperture, thereby exciting multiple orthogonal spatial modes. In the following section, we present a DFT-based decomposition of \ac{NF} beams to analyze the resulting \ac{EDoF}.
\subsection{Representation of NF Signal Using DFT Beams}
A \ac{BS} equipped with $M$ antennas can transmit up to $M$ spatially orthogonal beams, each associated with a discrete angular direction $\sin(\theta_m)\in[-1,1]$, represented by the columns of a \ac{DFT} codebook. For a \ac{FF} \ac{UE}, the correlation $\mathbf{a}^\mathsf{H}(\theta)\mathbf{a}(\theta_m)$ is maximized only when the user’s direction $\theta$ matches $\theta_m$, and ideally becomes negligible for all $\theta_m\neq\theta$. In the \ac{NF}, however, a \ac{UE} exhibits non-negligible correlation with multiple adjacent \ac{DFT} beams, revealing the presence of additional spatial \ac{DoF}. This behavior arises because far-field propagation is governed by planar wavefronts, whereas near-field propagation results in spherical wavefronts that can be expressed as a superposition of multiple angular (planar) components. Motivated by this observation, we model a \ac{NF} beam as a weighted combination of \ac{DFT} beams. The number of \ac{DFT} beams with high gain reflects the available spatial \ac{DoF}.

The \ac{EDoF} of a \ac{MIMO} system may also be interpreted as the number of transmit beams that can be uniquely mapped to distinct receive beams. The number of such unique pairs is upper bounded by the smaller of the numbers of transmit and receive beams. In principle, one could decompose the beams at both the transmitter and receiver to explicitly count these unique beam pairs. However, identifying the number of unique pairs is a complex task. To simplify the analysis, we project only the transmit-side \ac{NF} channel onto the angular domain using the \ac{DFT}. The number of DFT beams exhibiting high gain in this domain provides an estimate of the transmit-side \ac{EDoF}. The overall MIMO \ac{EDoF} is then computed by scaling this transmit-side estimate with the effective aperture length $D_r$ at the receiver. 

\begin{figure}[t]
\centering
    \includegraphics[width=.6\linewidth]{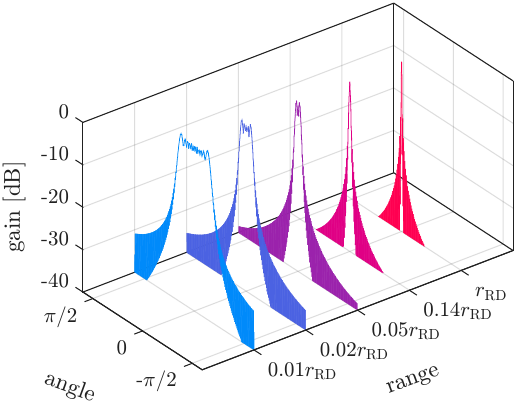}
    \caption{Angular spread (EDoF) at different ranges.}
    \setlength{\belowcaptionskip}{-10pt}
    \label{fig_EDoF_range}
    \end{figure}
\begin{figure}[t]    \centering
    \includegraphics[width=.85\linewidth]{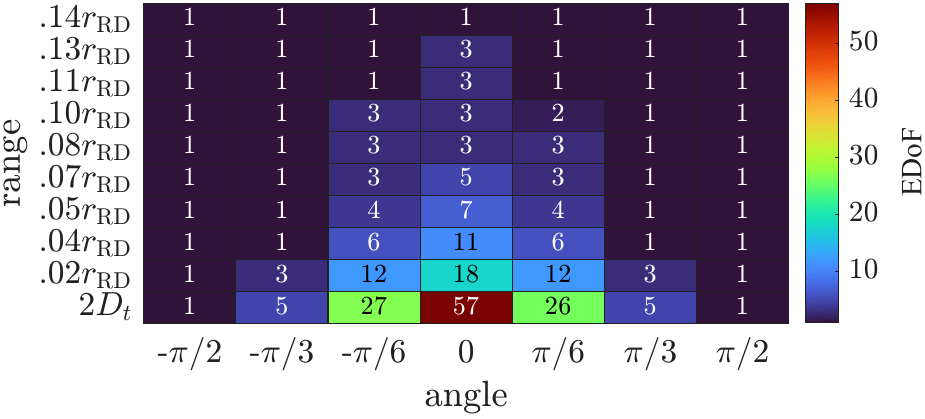}
    \setlength{\belowcaptionskip}{-20pt}
\setlength{\abovecaptionskip}{0pt}
    \caption{\ac{EDoF} for the transmitter at different ranges and angles. We set $M=256$, $f_c = \unit[29]{GHz}$, $D_t=\unit[1.3]{m}$, $\RD=\unit[336]{m}$.}
    \label{fig_EDoF_SIMO_table}
\end{figure}
We consider a \ac{NF} transmit beam focused at the location $(\theta,r)$. The \ac{NF} array response vector approximated based on Fresnel approximation is \cite{10934779}
\begin{equation}
\mathbf{b}_m(\theta,r) \approx 
\frac{1}{\sqrt{M}}
\, e^{ -j\frac{2\pi}{\lambda}
\left(
    m d \sin(\theta)
    - \frac{m^{2} d^{2} \cos^{2}(\theta)}{2r}
\right) },
\label{eq:b_n}
\end{equation}
while the \ac{FF} steering vector is obtained by omitting the quadratic phase term as
\begin{equation}
\mathbf{a}_m(\theta) \approx 
\frac{1}{\sqrt{M}}
\, e^{ -j\frac{2\pi}{\lambda}
    m d \sin(\theta)
}.
\label{eq:a_n}
\end{equation}
In the following, we derive the \ac{DFT} coefficients corresponding to the transmit-side \ac{NF} array response vector in \eqref{eq:b_n}. The resulting gain $\mathcal{G}$ observed at the \ac{NF} \ac{UE} from each \ac{DFT} beam directed toward $\theta_m$ is expressed as
\begin{equation}\small 
\begin{aligned}
&\mathcal{G}(\theta,r;\theta_m) ={\left| { \mathbf{b} ^{\mathsf{H}} (\theta,r) \mathbf{a} (\theta_m)} \right|}^2, \quad \sin(\theta_m)\in[-1,1]\\
&\approx
\tfrac{1}{M^2}\left|\sum_{-M/2}^{M/2} e^{j\tfrac{2\pi}{\lambda}\{md\sin(\theta)- \frac{1}{2r}m^2d^2\cos^2(\theta)\}-j\tfrac{2\pi md\sin\theta_m}{\lambda} } \right|^2,\\
&\stackrel{(c_1)}{=}
\tfrac{1}{M^2}\left|\sum_{-M/2}^{M} e^{-j\pi \{ m^2(\tfrac{d\cos^{2}\theta}{2r}) - m(\sin\theta - \sin\theta_m)\} } \right|^2,
\label{eqn9}
\end{aligned}
\end{equation}
\normalsize
where ($c_1$) is simplified assuming $d=\lambda/2$. We further simplify the array gain function in \eqref{eqn9} in terms of Fresnel functions $\mathcal{C(\cdot)}$ and $\mathcal{S(\cdot)}$, whose arguments depend on the UE location ($\theta,r$) and DFT beam angle $\theta_m$. Accordingly, the gain function observed at the \ac{NF} \ac{UE} under illumination from orthogonal \ac{DFT} beams can be approximated as \cite{11160729}
\begin{equation}
\begin{aligned}
\mathcal{G}(\theta,r;\theta_m) 
&\approx
\left|\frac{\overline{\mathcal{C}}\left(\gamma_{1}, \gamma_{2}\right)+j \overline{\mathcal{S}}\left(\gamma_{1}, \gamma_{2}\right)}{2 \gamma_{2}}\right|^2, 
\label{eqn10}
\end{aligned}
\end{equation}
$\overline{\mathcal{C}}\left(\gamma_{1}, \gamma_{2}\right) \equiv \mathcal{C}\left(\gamma_{1}+\gamma_{2}\right)-\mathcal{C}\left(\gamma_{1}-\gamma_{2}\right)$ and $\overline{\mathcal{S}}\left(\gamma_{1}, \gamma_{2}\right) \equiv$
 $\mathcal{S}\left(\gamma_{1}+\gamma_{2}\right)-\mathcal{S}\left(\gamma_{1}-\gamma_{2}\right)$, where $\gamma_{1}=\sqrt{\frac{r}{d\cos^{2}\theta}}(\sin\theta_m-\sin\theta) $ and $\gamma_{2}=\frac{M}{2} \sqrt{\frac{d\cos^{2}\theta}{r}}$. To characterize the \ac{EDoF}, we adopt the $\unit[3]{dB}$ threshold criterion. We count the angular directions $\theta_m$ with the normalized gain above $0.5$ and term it as the \ac{EDoF} for the transmitter given by
\begin{equation}
\mathsf{EDoF}_{2}(\theta,r) 
= \sum_{m}
\mathbf{1}\!\left\{
\mathcal{G}(\theta,r;\theta_m)  \geq \tfrac{1}{2}
\right\}.
\label{eqn_EDoF2}
\end{equation}
where $\mathbf{1}(\cdot)$ denotes the indicator function. We use the above expression to evaluate the \ac{EDoF} in our measurements. 

To gain further insight, we plot the gain in \eqref{eqn10} across different ranges, as illustrated in Fig.~\ref{fig_EDoF_range}. At the Rayleigh distance for \ac{MISO} given by $\RD = \tfrac{2D_t^2}{\lambda}$, which represents the classical boundary between the \ac{NF} and \ac{FF} regions, \eqref{eqn10} yields a single peak value for $\mathcal{G}$. In contrast, at shorter ranges, multiple peaks emerge around the \ac{UE} angle forming an angular spread, indicating an increased \ac{EDoF}. It is important to highlight that Fig.~\ref{fig_EDoF_range} can also be obtained by applying the $\operatorname{FFT}$ to the \ac{NF} channel in \eqref{eq:b_n} and plotting the resulting spectrum. Fig.~\ref{fig_EDoF_SIMO_table} presents the \ac{EDoF} computed via \eqref{eqn_EDoF2} at various range and angle values. The maximum \ac{EDoF} is achieved at the boresight direction and at a range equal to $2D_t$, where it reaches a value of $57$. This maximum \ac{EDoF} is significantly smaller than the total number of antenna elements, which is $M = 256$. Moreover, the minimum \ac{EDoF} of $1$ is observed for $r < \tfrac{\RD \cos^2\theta}{7}$ \cite{10934779}, which corresponds to the beam-focusing limit of a \ac{ULA}. This limit is seven times smaller than the Rayleigh distance at boresight. Additionally, for a fixed range, the \ac{EDoF} attains its maximum at boresight and gradually decrease toward the endfire directions.

The \ac{EDoF} in \eqref{eqn_EDoF2} represents the spatial \acp{DoF} offered solely by the transmit aperture, without accounting for the receiver. In a point-to-point \ac{MIMO} system, the net \ac{EDoF} is additionally influenced by the receive aperture. To incorporate the receive aperture length $D_r$, we combine \eqref{eqn_EDoF1} and \eqref{eqn_EDoF2} to obtain the following scaled distance:
\begin{equation}
    \hat{r} = \frac{D_r D_t}{\lambda \, \text{EDoF}_{2}}\cos\varphi_t \cos\varphi_r.
    \label{eqn_rescaled_distance}
\end{equation}
In recent studies, the Rayleigh distance for \ac{MIMO} systems has been defined based on the maximum allowable phase error (typically $\tfrac{\pi}{8}$) across the antenna aperture when approximating a spherical wave as a planar wave. However, this definition does not directly capture the system's ability to support independent spatial streams. As a result, the conventional Rayleigh distance does not reliably separate \ac{NF} and \ac{FF} regions. To address this limitation, we propose an alternative metric \ac{EMRD} that defines \ac{NF} boundaries in terms of \ac{EDoF}. Specifically, we define the \ac{EMRD}, denoted by $\MRD$, as the distance at which the \ac{EDoF} drops to one in a \ac{MIMO} system. By setting $\text{EDoF}_{2} = 1$ in \eqref{eqn_rescaled_distance}, this metric can be expressed as
\begin{equation}
    \MRD = \frac{D_r D_t}{\lambda}\cos\varphi_t \cos\varphi_r.
    \label{eqn_MRD}
\end{equation}
It is also desirable to derive a closed-form expression for the distance at which the spatial multiplexing is maximized, i.e., when the \ac{EDoF} attains its largest value. We refer to this distance as \ac{MSD}. This metric is obtained by replacing $\text{EDoF}_{2}$ in \eqref{eqn_rescaled_distance} with $V=\max\{M,N\}$, yielding
\begin{align}
    \MSD 
    &= \frac{D_r D_t}{\lambda\, V}\cos\varphi_t \cos\varphi_r, \quad V=\max\{M,N\}
    \label{eqn_MSD_general} \\[4pt]
    &= \frac{D_r}{2}\cos\varphi_t \cos\varphi_r,
    \quad \ M > N,
    \label{eqn_MSD_simplified}
\end{align}
where \eqref{eqn_MSD_simplified} follows by assuming $M > N$, substituting the aperture length $D_t \approx M d$, and applying the half-wavelength spacing condition $d = \lambda/2$ into \eqref{eqn_MSD_general}. Furthermore, \eqref{eqn_MSD_simplified} can also be derived by computing the inner product between the columns of $\mathbf{H}$ and enforcing orthogonality. Please refer to Appendix~\ref{Appendix_A} for the derivation, which is inspired by \cite{1424539}.

\section{Experimental Setup and Results}\label{sec_IV}
\begin{figure}[t]
\centering
\includegraphics[width=0.9\columnwidth]{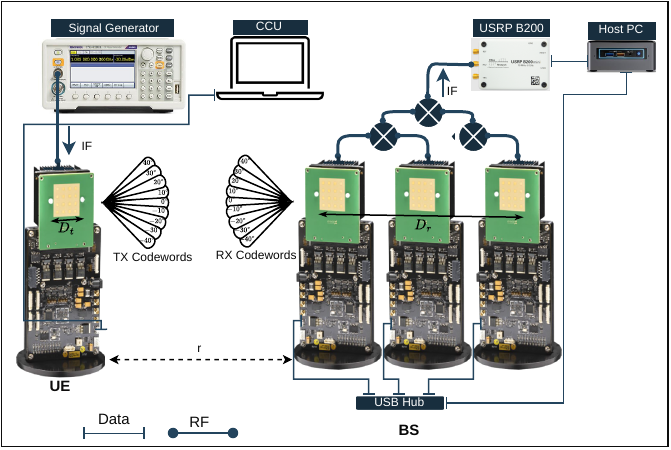}
\caption{Experimental setup for measuring the \ac{EDoF} in the \ac{NF} region.}
\setlength{\belowcaptionskip}{-20pt}
\setlength{\abovecaptionskip}{0pt}
\label{fig_experimental_setup}
\end{figure}

In this section, we describe the experimental setup shown in Fig.~\ref{fig_experimental_setup} and compare the measured results with the theoretical \ac{EDoF} to validate the analysis. The corresponding experimental parameters are summarized in Table~\ref{table:1}.
\subsection{Measurement Equipment and Configuration}
\subsubsection{BS}
We employ the EVK02004 phased-array, which incorporates a $4\times 4$ \ac{USA} of integrated patch elements. It features an in-built up/down-conversion stage that accepts \ac{IF} signals in the FR1 band and upconverts them to the $24$--$29.5\,\text{GHz}$ \ac{mmWave} band. It supports two-dimensional electronic beam steering over $\pm 40^\circ$ in both azimuth and elevation, with a beamwidth of $20^\circ$ in each dimension. In our experimental setup, we steer beams only along the azimuth axis and model the \ac{USA} as a \ac{ULA}. We integrate three phased arrays to extend the \ac{NF} communication distance. The aperture length of a single array along azimuth is $\unit[2.75]{cm}$, and it increases to $D_r=\unit[30]{cm}$ for the integrated array configuration. The \ac{RF} outputs of the three arrays are combined using an \ac{RF} power combiner and subsequently fed into the receiver port of a USRP~B200.

\begin{table}[t]
\caption{Experimental Parameters}
\centering
\begin{tabular}{ |c|c|c|c| }
\hline
\textbf{Parameter} & \textbf{Value} & \textbf{Parameter} & \textbf{Value} \\ [0.20ex]
\hline
Carrier Frequency & 29 GHz & IF Frequency & 5 GHz \\ [0.20ex]
\hline
Transmit Aperture ($D_t$) & 2.75 cm & Receive Aperture ($D_r$) & 30 cm \\ [0.20ex]
\hline
Beamwidth ($\theta_{3\text{dB}})$ & $20^\circ$ & Tx/Rx codewords & 9\\ [0.20ex]
\hline
\end{tabular}
\label{table:1}
\end{table}

\subsubsection{UE}
The \ac{UE} is equipped with a single EVK02004 phased-array module, which serves as the transmitter. A signal generator supplies a \ac{IF} tone at \unit[5]{GHz} to the module, which then upconverts the signal to carrier frequency of  $f_c=$ \unit[29]{GHz}.
\subsubsection{Central Control Unit (CCU)}
We interface with the EVK modules using two separate host computers, each connected to its corresponding EVK unit via a USB. The host computer on the \ac{UE} side functions as the \ac{CCU} and communicates with the main host PC via wireless TCP/IP and Bluetooth links. The \ac{CCU} serves as the central controller of the testbed, coordinating all hardware components, managing control signaling, and ensuring system-wide synchronization and signal processing. 
\subsubsection{DFT Codebook}
Each antenna element of the phased array is equipped with a 2-bit phase shifter, enabling dynamic control of the beam pattern. Both the \ac{BS} and \ac{UE} are configured with identical DFT codebooks that scan the azimuth angle from $-40^\circ$ to $40^\circ$ in steps of $10^\circ$. Since the beamwidth is $20^\circ$, this results in an oversampled DFT codebook consisting of nine codewords. The \ac{UE} sequentially transmits nine beams for each receive codeword at the \ac{BS}. For every Tx–Rx beam pair, we record the \ac{RSSI} using the USRP and forward the measurements to the host computer.

\subsection{Results}
\begin{figure}[t]
\centering
\includegraphics[width = .9\columnwidth]
{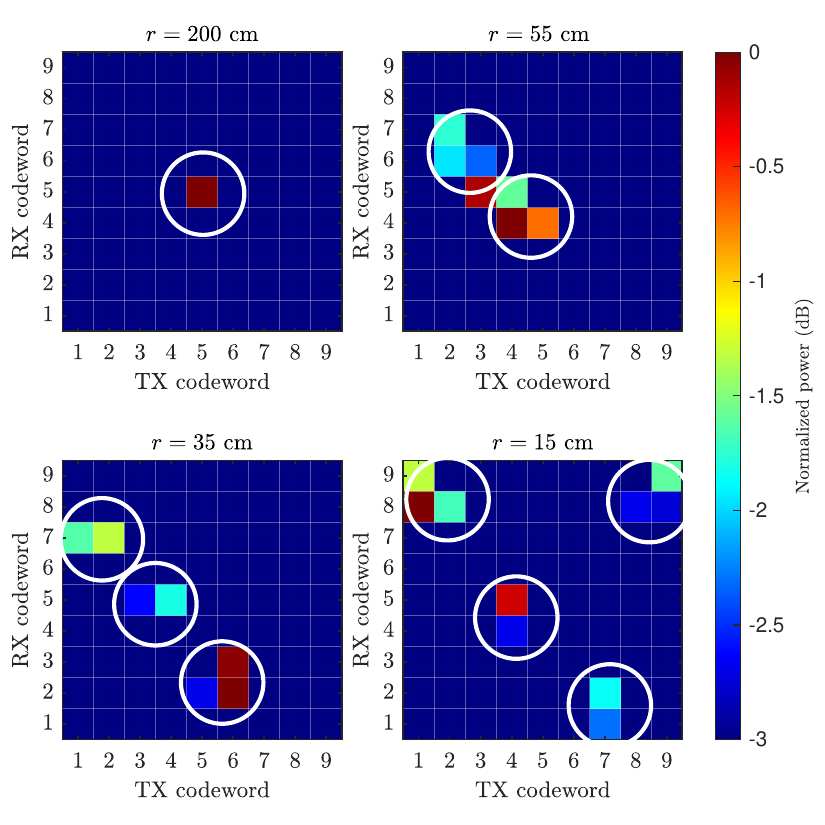}
\setlength{\belowcaptionskip}{-10pt}
\caption{Heatmap of the RSSI matrix for selected range values.}
\label{fig_RSSI_matrix}
\end{figure}

We align the \ac{BS} and \ac{UE} along the boresight direction ($\varphi_t=\varphi_r=0^\circ$) and vary the separation distance $r$ from \unit[200]{cm} down to \unit[15]{cm}. Fig.~\ref{fig_RSSI_matrix} presents the \ac{RSSI} heatmaps for selected distance values. For each Rx codeword, the Tx sequentially scans all nine Rx codewords before switching to the next Tx codeword. Repeating this over all nine Tx codewords produces a $9\times 9$ \ac{RSSI} matrix for every measured distance. The received power values are then normalized and clipped to a \unit[3]{dB} dynamic range, consistent with \eqref{eqn_EDoF2}. At $r=\unit[200]{cm}$, a single dominant peak appears, indicating an \ac{EDoF} of one. As the distance decreases, multiple peaks progressively emerge in the \ac{RSSI} maps, revealing the onset of \ac{NF} effects. For example, at $r=\unit[55]{cm}$, the Tx codewords span four bins (bins $2$–$5$) and the Rx codewords span four bins (bins $4$–$7$). Since the \ac{DFT} codebook is oversampled by a factor of two, these peaks form two clear clusters, corresponding to an \ac{EDoF} of two. At $r=\unit[35]{cm}$, three clusters are observed, while at $r=\unit[15]{cm}$, four clusters appear, yielding \acp{EDoF} of three and four, respectively. Furthermore, the inter-cluster separation increases as $r$ decreases, indicating reduced overlap and stronger orthogonality. Based on these measurements, we design heuristic rules to estimate the \ac{EDoF} directly from the raw \ac{RSSI} matrices. Peak locations are first extracted using a \unit[3]{dB} threshold. Clusters are then formed using two criteria: (i) peaks in the same row or column are grouped together, and (ii) peaks separated by only a single row or column are merged into one cluster. Each cluster is subsequently represented by the rounded centroid of its peak coordinates, and the \ac{EDoF} is obtained as the number of resulting clusters.

\begin{figure}[t]
\centering
\includegraphics[width = 1\columnwidth]
{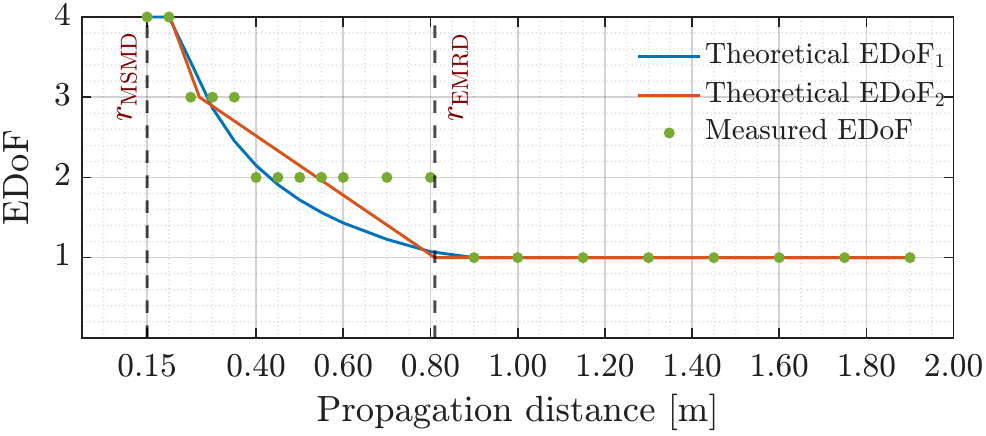}
\setlength{\belowcaptionskip}{-20pt}
\caption{Comparison of theoretical and experimental \ac{EDoF}.}
\label{fig_EDoF_com}
\end{figure}

We plot the resulting \ac{EDoF} obtained from measurements as a function of distance in Fig.~\ref{fig_EDoF_com}. These results are compared against the theoretical expressions derived in Section~\ref{sec-III}. For the given aperture sizes $D_t$ and $D_r$, the \ac{EMRD} based on \eqref{eqn_MRD} is
$ \MRD = \frac{D_t D_r}{\lambda} = \frac{0.30 \times 0.02}{0.0075} = 0.81~\text{m} $.
Similarly, the \ac{MSD} from \eqref{eqn_MSD_simplified} is
$ \MSD = \frac{D_r}{2} = 0.15~\text{m} $.
As shown in Fig.~\ref{fig_EDoF_com}, the measured \ac{EDoF} reaches its maximum value at $r = \unit[15]{cm}$, which agrees with the predicted $\MSD$. As the link distance increases, the \ac{EDoF} decrease and eventually reduces to one as the distance approaches the $\MRD$.  

Achieving practical \ac{EMRD} values requires very large transmit and receive apertures, which can be costly in hardware. A feasible alternative is to deploy widely spaced subarrays at the \ac{BS}, effectively enlarging the aperture. In our measurements, the distributed phased arrays emulate a large effective aperture while reducing hardware complexity. The results confirm that higher \ac{EDoF} can be realized in \ac{LoS} scenarios using such widely spaced subarrays.

\section{Conclusion} \label{sec_VI}
In this letter, we developed a framework to characterize the \ac{EDoF} in \ac{ULA}-based \ac{MIMO} systems. Analytical expressions and experimental measurements using modular phased arrays demonstrate that the \ac{NF} can support significantly higher \ac{EDoF} compared to the \ac{FF}. These results provide valuable insights for designing high-capacity, low-complexity \ac{MIMO} systems in the radiative \ac{NF}.

\bibliographystyle{IEEEtran}
\bibliography{Bibliography/IEEEabrv,Bibliography/my2bib}

\appendix
\section{Appendix } 
\label{Appendix_A}
To derive the \ac{MSD} from the orthogonality condition of $\mathbf{H}$, we compute the inner product between the received vectors corresponding to two distinct transmit antennas, using the channel coefficients in \eqref{eqn_A1}, and enforce this inner product to be zero:
\begin{align}
\langle \mathbf{h}_{k,n} , \mathbf{h}_{l,n} \rangle
&= \sum_{n=0}^{N-1} 
    \exp\!\left( j\frac{2\pi}{\lambda}( r_{k,n} - r_{l,n} ) \right) \nonumber \\
&= \sum_{n=0}^{N-1}
    \exp\!\left( j \frac{2\pi\, d^2 \cos\varphi_r \cos\varphi_t}{\lambda r}\, (k-l)\, n \right) \nonumber  \\
&= 
\frac{\sin\!\left( \pi \frac{d^2}{\lambda r} \cos\theta_r \cos\varphi_t (k-l) N \right)}
     {\sin\!\left( \pi \frac{d^2}{\lambda r} \cos\varphi_r \cos\varphi_t (k-l) \right)}
     = 0, \nonumber\\                                       
\Rightarrow\quad 
r 
&= \frac{Nd^2}{\lambda} \cos\varphi_r \cos\varphi_t = \frac{D_r}{2} \cos\varphi_r \cos\varphi_t 
\label{eq:spacing_result}
\end{align}

\end{document}